# IAA-PDC-21-11-37
# Precautionary Planetary Defence

**Aaron C. Boley**[(1)], **Michael Byers**[(2)]
[(1)(2)] *Outer Space Institute*
*University of British Columbia*
*325-6224 Agricultural Road*
*Vancouver, BC V6T 1Z1 Canada*
*+1-604-827-2641*
*aaron.boley@ubc.ca*

*Introduction:* The question of whether to attempt deflections during planetary defence emergencies has been subject to considerable decision-making analysis (Schmidt 2018; SMPAG Ad-Hoc Working Group on Legal Issues 2020). Hypothetical situations usually involve a newly discovered asteroid with a high impact probability on a set timescale. This paper addresses two further complexities: (1) limiting missions to an asteroid due to the risk of a human-caused Earth impact; and (2) active management of asteroids to place them in "safe harbours", even when impact risks are otherwise below "decision to act" thresholds. We use Apophis as a case study, and address the two complexities in turn.

*Showing restraint:* Apophis's 2029 approach of approximately 38010 km[1] from the geocentre presents rare science opportunities for studying how very close encounters can alter asteroid spin states and lead to resurfacing, as well as for potentially probing the asteroid's interior structure. The close approach also offers public outreach opportunities, owing to Apophis being momentarily closer to us than geosynchronous satellites and the feasibility for some portions of the world's population to see the asteroid with an unaided eye. Science interest is motivating many different teams to propose a variety of missions, including rendezvous and flybys (for example, see the T-9 Apophis workshop program[2]). Some proposed missions are primarily technology demonstrations.

Although these opportunities are exciting, we also know that Apophis's 2029 b-plane has multiple keyhole complexes (e.g., Farnocchia et al. 2013). The current orbital solution, supplemented by recent Goldstone/GBT radar measurements, is sufficiently constrained that the passage of Apophis through a keyhole is improbable (CNEOS 2021). However, the extensive interest in visiting the asteroid before the 2029 close approach leads us to ask, in a general sense, to what degree might restraint be prudent?

As discussed by Chesley and Farnocchia (2021), if a mission to an asteroid with a rich set of keyholes, like Apophis, goes awry and unintentionally collides with the asteroid, there is a risk that this will create a future impact emergency. The publicity associated with the asteroid's close approach could also prompt non-state actors to launch their own missions as technology demonstrations and/or profile-raising exercises, much like the infamous Tesla launch by SpaceX.

Adding to these considerations is a potential traffic management problem should multiple independent missions be launched.

For these reasons, it makes sense to apply the precautionary principle, which is well-established in other areas of international relations (Freestone & Hey 1996), to this domain. Doing so will not preclude missions to asteroids such as Apophis, but it will demand a high level of coordination among space actors. It will also require that some missions be modified and, in extreme cases, highly limited despite the scientific or technology demonstration benefits that might otherwise be gained. This immediately leads one to ask: Who should be making these decisions?

*Decisions and Coordination:* The Space Mission Planning Advisory Group (SMPAG) is composed of representatives from 18 space agencies and aims to "develop cooperative activities among its members and to build consensus on recommendations for planetary defense measures" (SMPAG 2019). However, SMPAG does not have decision-making authority. Even SMPAG members do not require permission from the group to carry out a mission. Rather, national governments make

---

[1] See JPL Small-Body Database Browser, Apophis close approach tables.

[2] Conference abstracts at: https://www.hou.usra.edu/meetings/apophis2020/

the decision whether to proceed with a mission, either on their own or in cooperation with others.

Provided that past levels of cooperation among major spacefaring states are maintained, SMPAG provides a working framework for collectively determining whether any limitations are needed on a proposed planetary defence mission. However, there are a number of ways in which this cooperation might break down.

US plans for the Moon, which have already seen contracts signed for the purchase of regolith from private companies with the explicit goal of establishing legal precedents for commercial space mining, have recently created tensions within the existing cooperative framework (Boley & Byers 2020). In particular, Russia and China have rejected the US-led Artemis Accords and announced plans to develop their own model for lunar governance and infrastructure (AP 2021). Distrust among the major spacefaring states could also increase if the US continues with plans to send military spacecraft into cis-lunar space (Hitchens 2021).

A completely new dimension to decision-making comes with NewSpace, with some private actors expected to possess advanced exploration capabilities well before 2029. In 2019, SpaceIL became the first non-state entity to place a spacecraft on the Moon, albeit via a hard landing. SpaceX is already flight-testing Starship, a reusable spacecraft for Earth orbit, the Moon, and Mars. Multiple tourism ventures are underway, with trips to the ISS and around the Moon planned for the next 2-3 years. One or more of these increasingly capable space actors may wish to use Apophis or other asteroids for their own purposes. The prospect of eventual asteroid mining adds yet another dimension, as this could be a benefit or a risk to planetary defence depending on the degree to which companies share information, some of which they may consider proprietary.

Although the Outer Space Treaty makes the "launching state" responsible for their actions, non-state actors might have different approaches to scientific uncertainty and risk. They might not engage, or be required to engage, in the same level of coordination as national agencies do through SMPAG. Nor are all national regulatory frameworks necessarily prepared for a much higher level (and volume) of commercial space activity.

National regulators should be strongly encouraged to take planetary defence considerations into account when issuing launch licenses to non-state actors, including adopting practices that require both the non-state actor and the regulator to consult with SMPAG.

One body does have ultimate say and could in principle limit missions to an asteroid: The UN Security Council. However, a Security Council resolution is a heavy-handed solution to what should be an avoidable problem. Moreover, to be adopted, such resolutions must be supported by at least nine of the 15 members of the Council with no vetoes cast by any of the five permanent members (China, Russia, US, UK, France). Yet the Security Council could usefully become engaged in a preparatory manner, for instance, by adopting a resolution on planetary defence matters in general. Such a resolution could require any state planning or licensing a mission to an asteroid to consult with SMPAG and satisfy any concerns that it might have.

**Active Management:** Precautionary planetary defence is not just about showing restraint; it includes the active management of asteroid orbits. Essentially, there might be situations where the redirection of an asteroid is warranted even if it does not yet pose a risk or if the risk is unknown. Such active "shepherding" would ideally be conducted with a gravity tractor to ensure minimal interference with the asteroid. Yeomans et al. (2009) explored this in the context of Apophis when there was still some worry that it might pass through a 2029 keyhole. One of the points in that paper – of finding so-called "safe harbours" – remains salient as one can always ask whether there is an accessible orbit that not only misses keyholes but also minimizes the long-term risk posed by a given asteroid. We again use Apophis to illustrate our point, although this discussion could be applicable to multiple asteroids, including Bennu. First, however, there needs to be some metric for defining the safest accessible harbour. We do not propose the metric here, but sketch some considerations.

We start by showing in Figure 1 the close approach profile of Apophis on the 2029 b-plane, which is the minimum distance that Apophis has with Earth after the 2029 flyby over the next 100 yr. The minimum distances are shown with respect to the $\zeta$ coordinate. Parameter space is explored by taking the current orbital solution and perturbing Apophis at the start of the simulation along or against track using systematically increasing $\Delta v$'s. Simulations were run using a modified version of Rebound/X (Rein & Spiegel 2015; Tamayo et al. 2019) and included GR, Earth's J2 and J4 components, and perturbations from the list of asteroids given in Farnocchia et al. (2013).

From this figure we can see that Apophis is close to a downward spike in the minimum distance profile (corresponding to the 2116 encounter), potentially dropping below a 30 Earth radii minimum distance. Importantly, it is not an impact keyhole, so the location is safe despite the potential for a future close encounter. At slightly higher $\zeta$, Apophis can be kept farther from Earth over the next 100 yr than at its current nominal location. However, such a change would place the asteroid closer to a keyhole complex. At lower $\zeta$, a small "hill" exists that is free from close encounters and known keyholes. Which is the safest harbour?

In the above case, one might argue against moving Apophis to higher $\zeta$ on the grounds that is closer to a keyhole. A response to this concern might be that a rendezvous with a gravity tractor should enable a precise orbit to be determined, in which case the shepherding could always be reassessed. This might include aborting the mission if the orbital uncertainty remains too high.

Moving to lower ζ does not raise the same concern, though it also does not lead to a much better situation than that provided by the current orbit.

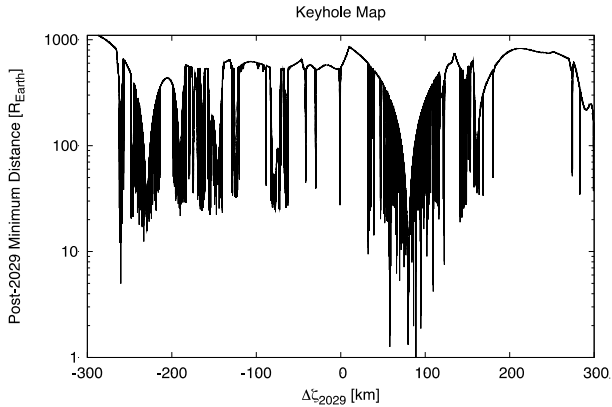

*Figure 1: Keyhole map using Rebound/X simulations of Apophis. The current Horizons orbit solution (soln. ref. JPL211, epoch 2021-April-7.0) is used for initial conditions. The map is produced by giving virtual Apophis particles an instantaneous perturbation along or against track at the start of the simulation (at epoch), with the magnitude progressively increasing. The nominal solution is near a non-hazardous downward spike (2116). B-plane locations to the left (small hill) and right (cusp) might offer safer harbours. We use $\zeta_{2029} = 47362.5$ km as the nominal b-plane location for $\Delta\zeta_{2029}$.*

While again we do not know the exact metric that should be used, qualitatively, if we have the means to substantially increase the long-term close-approach distance of a potentially hazardous asteroid then we should consider doing so. Otherwise, the asteroid should not be perturbed. In our thought experiment with Apophis, this reasoning might argue for nudging the asteroid slightly to higher ζ.

We might further repose the question by imagining what our reaction would be if Apophis were in a narrow, safe region within the keyhole complex at, for example, $\Delta\zeta_{2029} = 75$ km. Would this motivate deflection to lower ζ to find a safer harbour?

While considering these questions, we must acknowledge that the precautionary principle could support an argument against active management because such an approach might create new risks. For example, if a failure happened while tractoring, a given asteroid could be dropped into a keyhole (Yeomans et al. 2009). Thus, a decision to actively manage an asteroid into a safe harbour should only be taken after peer-reviewed scientific assessment, full international collaboration, and broad agreement.

**Management with Gravity Tractors:** A detailed model for a gravity tractor is given in Yeomans et al. (2009), demonstrating multiple deflection scenarios that are feasible with an acceleration of approximately $10^{-12}$ m s$^{-2}$ exerted by a 1000 kg spacecraft.

Returning to the minimum post-2029 flyby distance for Apophis, we show in Figure 2 possible deflections as an illustrative example of a safest harbour search. More specifically, we ran a series of gravity tractor simulations to explore conditions that could place Apophis into the cusp at slightly higher ζ. We used the same setup as we did for Figure 1, but applied an acceleration along or against track for different time intervals, with durations lasting 6 months or 1 yr. Without assuming a specific tractor design, we used an acceleration of $10^{-12}$ m s$^{-2}$ for the 1 yr duration tractor and $10^{-11}$ m s$^{-2}$ for the 6-month tractor (see figure caption for details). The results show that redirecting the asteroid to a potentially safer harbour is possible with a gravity tractor, even on short timescales. The low-impulse case can place the asteroid near the cusp, provided tractoring starts around 2026 or earlier. The high-thrust situation can move the asteroid into the cusp or hill if it begins at least one year prior to the 2029 encounter. Again, this is not to argue that we should attempt such a manoeuvre, but rather, highlights possibilities.

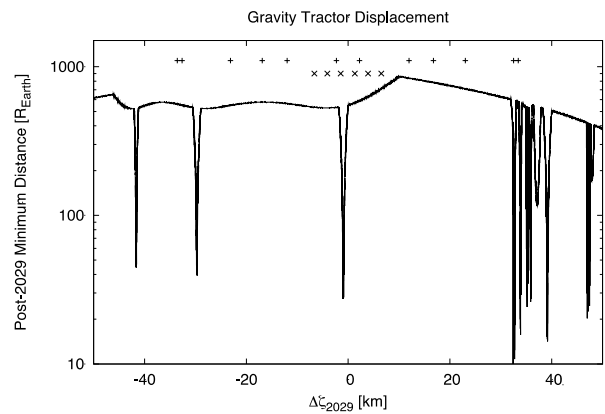

*Figure 2: Zoom in of the keyhole map, centred on the nominal solution. The symbols along the top of the figure represent the locations after different gravity tractor manoeuvres. The crosses (x) correspond to an acceleration of $10^{-12}$ m s$^{-2}$ for three different intervals (along and against track). These intervals are approximately (outward to inward) 7 April 2026-2027, 2027-2028, and 2028-2029. The plusses (+) correspond to $10^{-11}$ m s$^{-2}$. Due to the higher thrust, only 6-month intervals are considered. Moving outward to inward, the starting dates are approximately 7 April 2026, 7 October 2026, 7 April 2027, 7 October 2027, 7 April 2028, and 7 October 2028. Note that pericentre occurs near the end of July.*

In closing, we note that SpaceX's Starship is about 100 t empty, and that it is fully automated and reusable. Designed to transport and land cargo and people on the Moon and Mars, a version of Starship could be reconfigured as a highly effective and reusable gravity tractor. Given the rapid pace of development and flight testing currently underway, Starship provides a potential option for performing a gravity tractor manoeuvre before or after the Apophis 2029 flyby.

**Summary:** A key aspect of planetary defence is risk analysis and prevention. This should include precaution with respect to asteroid missions by state and non-state actors, consideration of the active management of some asteroid orbits, and governance innovations aimed at ensuring that international collaboration always takes place. The scope of planetary defence decision-making should thus be expanded beyond reactions to specific potential impacts. These changes are necessitated – and to some degree, made possible – by the emergence of a broader set of actors planning an increasing number of missions.

Apophis does not pose a threat at this time, and the discussion here should not be read as advocating active management specifically of it; nor are we advocating that pre-2029 encounter missions be severely restricted. Rather, Apophis's history and keyhole profile provide a working example for raising the issue of precautionary planetary defence and demonstrating – through calculations designed to show the availability of safe harbours – how it might operate in practice.